\documentclass[a4paper,12pt]{article}
\usepackage{graphicx,amssymb,bm,latexsym}
\usepackage{geometry}
\usepackage{epsf,colordvi}
\newcommand{\beq}{\begin{equation}}
\newcommand{\eeq}{\end{equation}}
\newcommand{\bea}{\begin{eqnarray}}
\newcommand{\ena}{\end{eqnarray}}
\newcommand{\vs}[1]{\vspace{#1 mm}}

\renewcommand{\a}{\alpha}

\newcommand{\e}{\epsilon}

\newcommand{\dsl}{\pa \kern-0.5em /}

\newcommand{\pa}{\partial}

\newcommand{\nn}{\nonumber\\}
\newcommand{\p}[1]{(\ref{#1})}

\begin{document}

\begin{titlepage}
\begin{flushright}
KU-TP 050 \\
\phantom{arXiv:yymm.nnnn}
\end{flushright}
\begin{center}
\vskip 3cm
{\Large\bf Extended black holes in strong gravitational waves}    \\
\vskip 10mm
{\large Li-Ming Cao$^a$, Oleg Evnin$^b$ and Nobuyoshi Ohta$^a$}
\vskip 7mm
{\em $^a$ Department of Physics, Kinki University\\
Higashi-Osaka, Osaka 577-8502, Japan}
\vskip 3mm
{\em $^b$ Institute of Theoretical Physics, Academia Sinica\\
Zh\=onggu\=anc\=un d\=ongl\`u 55, Beijing 100190, China}
\vskip 3mm
{\small\noindent  {\tt caolm@phys.kindai.ac.jp, eoe@itp.ac.cn, ohtan@phys.kindai.ac.jp}}
\end{center}
\vfill

\begin{center}
{\bf ABSTRACT}\vspace{3mm}
\end{center}
We describe a large class of solutions in pure gravity, dilaton gravity and supergravity
corresponding to extended higher-dimensional black holes with strong (non-linear)
gravitational waves propagating along their worldvolume. For pure gravity,
the extended black holes are higher-dimensional analogs of the point-like Schwarzschild black
hole in four dimensions. For supergravity, they are non-extremal $p$-branes.
The gravitational waves can be both space-filling and localized around the worldvolume
of the extended black holes. The solutions we present contain a large number of
arbitrary functions of the light-cone time describing the amplitudes of different
non-linear gravitational wave modes.

\vfill

\end{titlepage}

\section{Introduction}

Exact interactions of strongly non-linear objects are generally difficult to describe,
and the cases when an analytic treatment can be given are rare. In a gravitational
theory, such results give an exact description of strong gravitational effects.
Here, we shall
present solutions in pure gravity, dilaton gravity and supergravity corresponding to extended
black holes with strong gravitational waves propagating along their worldvolume. There are
not many known solutions for black holes in time-dependent (or cosmological) backgrounds
(examples include black holes in Friedmann-Robertson-Walker \cite{mcvittie,FRW} and
de-Sitter \cite{dS,Bousso:2002fq} spacetimes, as well as supergravity $p$-branes embedded
in dilaton cosmologies \cite{Maeda:2009zi}).
A specific feature of our solutions is that they contain a large number of arbitrary
functions of the light-cone time corresponding to the profiles of the various polarization
components of the strong gravitational waves involved.

Strong gravitational waves in flat space-time are known to be described by the metric
(see, e.g., Appendix A of \cite{Blau:2008bp})
\beq
ds^2=-2du\,dv+K_{ij}(u)x^ix^jdu^2+(dx^i)^2,
\label{brink}
\eeq
where $K_{ij}(u)$ represent the profiles of different polarization components of the wave.
In pure gravity, $K_{ij}(u)$ is constrained by $K_{ii}=0$, giving the same number
of polarizations as in linearized theory (a traceless symmetric tensor in
$D-2$ dimensions, with $D$ being the number of dimensions of space-time).
If a dilaton is present, $K_{ii}$ does not vanish and is related to the dilaton,
which gives an additional independent polarization component. The metric (\ref{brink})
is given in the so-called Brinkmann coordinates (which are not prone to coordinate
singularities and hence useful for global considerations). By a $u$-dependent
rescaling of $x^i$, it can be brought to the so-called Rosen form, in which the metric
only depends on $u$, making the planar nature of the wave front manifest.

In this paper, we shall consider the (considerably more complex) analogs of (\ref{brink}),
in which the strong gravitational wave propagates along an extended black hole.
The black hole can be an extended higher-dimensional Schwarzschild-like object in
pure gravity, or its dilaton gravity modification, or a non-extremal $p$-brane in
supergravity.
Because of the presence of the black hole, our solution will display two kinds of
non-linear modes: those approaching (\ref{brink}) far away from the black hole, and
those localized near the extended black hole worldvolume.

Simpler supersymmetric analogs of our present solutions have been previously considered in
a series of publications \cite{intersect}--\cite{MOTW1} with specific choices of the wave
profile (for some related literature, see \cite{blackstring}--\cite{Narayan}).
Extremal supersymmetric solutions with an arbitrary gravitational wave profile were
constructed in \cite{CDEG}, and solutions with an arbitrary profile featuring intersecting
$p$-branes were constructed in \cite{MOTW2}. Here, we generalize the previous derivations
to the case of
non-extremal non-supersymmetric extended black holes along the lines suggested in \cite{CDEG}.

The paper is organized in a straightforward manner: we first present the general equations
of motion and discuss the structure that permits their thorough analysis. We then proceed
with the more computationally transparent case of dilaton gravity
(that includes pure gravity). Finally, we present the analysis of the non-extremal
supergravity $p$-branes.

\section{Equations of motion}

We start with the low-energy effective action for the supergravity system
coupled to dilaton and $n_A$-form field strengths (pure gravity and dilaton gravity
are obtained straightforwardly by setting some of the fields to zero):
\bea
I = \frac{1}{16 \pi G_D} \int d^D x \sqrt{\mathstrut-g} \left[
 R - \frac12 (\pa \phi)^2 - \sum_{A=1}^m \frac{1}{2 n_A!} e^{a_A \phi}
 F_{n_A}^2 \right],
\label{action}
\ena
where $G_D$ is the Newton constant in $D$
dimensions and $g$ is the determinant of the metric. The last term
includes both RR and NS-NS field strengths, and $a_A = \frac12
(5-n_A)$ for RR field strength and $a_A = -1$ for NS-NS 3-form.
We put fermions and other background fields to be zero.

{}From the action (\ref{action}), one can derive the field equations
\bea
R_{\mu\nu} = \frac12 \pa_\mu
\phi \pa_\nu \phi + \sum_{A} \frac{1}{2 n_A!}
 e^{a_A \phi} \Biggl[ n_A \left( F_{n_A}^2 \right)_{\mu\nu}
 - \frac{n_A -1}{D-2} F_{n_A}^2 g_{\mu\nu} \Biggr],
\label{Einstein}
\ena
\vs{-5}
\bea
\Box \phi = \sum_{A} \frac{a_A}{2 n_A!}
e^{a_A \phi} F_{n_A}^2,
\label{dila}
\ena
\vs{-5}
\bea
\pa_{\mu_1} \left(
\sqrt{\mathstrut- g} e^{a_A \phi} F^{\mu_1 \cdots \mu_{n_A}} \right) = 0
\,,
\label{field}
\ena
where $F_{n_A}^2$ denotes $F_{\mu_1 \cdots \mu_{n_A}}F^{\mu_1 \cdots \mu_{n_A}}$
and $(F_{n_A}^2)_{\mu\nu}$ denotes $F_{\mu\rho\cdots\sigma}F_\nu^{~\rho\cdots\sigma}$.
The Bianchi identity for the form field is given by
\bea
\pa _{[\mu} F_{\mu_1 \cdots \mu_{n_A}]} =0.
\label{bianchi}
\ena

In this paper we consider a general $D$-dimensional theory (which includes the ten-dimensional supergravity case) and assume the following metric form:
\bea
ds_D^2 = e^{2A(u,r)} \left[-2dudv + K(u, r, y^\alpha) du^2\right] +
e^{2C(u,r)} (dy^\a)^2 + e^{2B(u,r)} \left(dr^2 + r^2 d\Omega_{ q+1}^2\right),
\label{met}
\ena
where the coordinates $u$, $v$ and $y^\a, (\a=1,\ldots, p-1)$
parameterize the $(p+1)$-dimensional worldvolume of the extended black hole,
and the remaining $(q+2)$ coordinates $r$ and angles are transverse to the brane
worldvolume, $p+ q+3=D$, $d\Omega_{ q+1}^2$ is the line element of the
$( q+1)$-dimensional sphere. If the $y^\alpha$-dependence of $K$
is removed in (\ref{met}), the metric represents the most general space-time
with rotational symmetries of $x^a$ and $y^\alpha$ and translational symmetries
of $v$ and $y^\alpha$ (with the $v$-translation being light-like). It turns out that
the $y^\alpha$-dependence can be added to $K$ with only minimal modifications to the
equations of motion (and it is useful for including $y^\alpha$-polarizations of the
space-filling plane waves). Such an ansatz has already been considered in \cite{MOTW1}
for supersymmetric solutions.
All the remaining components of the metric and the dilaton are assumed
to be functions of $u$ and $r$ only.

For the background RR-form, we take
\bea
F_{p+2} = E'(u,r) du\wedge dv \wedge dy^1 \wedge \cdots \wedge dy^{p-1}\wedge dr,\qquad
n=p+2,~~ a=\frac{3-p}{2}.
\ena
Throughout this paper, the prime and dot denote the derivatives with respect to
$r$ and $u$, respectively, and $a$, $b$ are angular coordinates.
We could also include magnetic background in the same form as the electric one.
The result gives basically the same set of equations with minor modifications.

The Einstein and dilaton equations are given as~\cite{MOTW1}
\bea
&& (p-1) (\ddot{C} - 2 \dot A\dot{C} +\dot{C}^2)
+( q+2)( \ddot B - 2 \dot A \dot B + \dot B^2)
+e^{2(A-B)} \Big[ KA''+\frac{1}{2}K'' \nn
\label{f1}
&& +\Big(KA'+\frac{1}{2} K'\Big)\Big( U'+\frac{ q+1}{r}\Big) \Big]
+\frac12 e^{2(A-C)}\partial_\alpha^2 K
=-\frac12 \dot\phi^2 + \frac{ q}{16} KS(E')^2 e^{2(A-B)}, \\
\label{f3}
&& {\dot A}' + (p-1) ( \dot{C}' - \dot{C} A'- C' \dot B
+ \dot{C} C')+ ( q+1) \dot B' - ( q+2) A' \dot B = -\frac12 \dot\phi \phi', \\
\label{f2}
&& A'' +  \Big( U'+\frac{ q+1}{r}\Big) A' = \frac{ q}{16} S(E')^2, \\
\label{f4}
&& C'' + \Big( U'+\frac{ q+1}{r}\Big) C' = \frac{ q}{16} S(E')^2, \\
&& U'' +B'' - \Big( 2A'+(p-1) C' - \frac{ q+1}{r}\Big)B' +2A'^2 +(p-1) C'^2 \nn
\label{f5}
&& \qquad = - \frac12\phi'^2+\frac{ q}{16} S (E')^2, \\
\label{f6}
&& B'' + \Big( U'+\frac{ q+1}{r}\Big)B'+ \frac{U'}{r}
= - \frac{p+1}{16} S (E')^2,\\
\label{f7}
&& e^{-U}r^{-( q+1)} (e^U r^{ q+1} \phi')' = \frac{p-3}{4} \e S (E')^2,\\
\label{f8}
&& (r^{ q+1} e^U SE' )'=(r^{ q+1} e^U SE' )^{\bm .}=0,
\ena
where we have defined
\bea
U \equiv 2A+(p-1) C+ q B, \qquad
S \equiv e^{\e (3-p)\phi/2 -2(2A+(p-1)C)},
\ena
and $\e=+1(-1)$ is for electric (magnetic) background.
Unlike in \cite{MOTW1}, we no longer constrain $U$
to depend only on $u$, but consider general solutions for our ansatz.

The general structure of the equations of motion has been already spelled out
in \cite{CDEG}: equations (\ref{f2})--(\ref{f6}) are exactly identical to those for
a $u$-independent problem, i.e., to the equations for an extended black hole without
any gravitational waves. Such equations have been extensively discussed in the literature
(a large class of solutions has been presented in~\cite{GKOC}). In our context,
we should take the most general solution to Eqs.~(\ref{f2})--(\ref{f6}), i.e., a static
extended black hole, and then promote all the integration constants to functions of $u$.
The resulting expressions should be substituted into Eq.~(\ref{f3}), which will impose some
constraints on the $u$-dependences of the arbitrary functions of $u$ (non-linear wave
amplitudes) contained in the solution of Eqs.~(\ref{f2})--(\ref{f6}). Finally, Eq.~(\ref{f1})
will determine $K$ (without introducing any further constraints). This structure
essentially reduces a large part of our problem
to constructing static extended black hole solutions, a thoroughly explored subject.

Implementing this program, we first learn from Eq.~\p{f8} that
\bea
\label{QSE'constant}
r^{ q+1} e^U SE'=c,
\ena
is a constant (related to the $p$-form charge). We find that Eq.~\p{f2} then gives
\bea
(e^U r^{ q+1}A')'=\frac{ q}{16}c E',
\ena
which can be immediately integrated to yield
\bea
A' e^U r^{ q+1} = \frac{ q}{16}cE + f_A(u),
\label{r1}
\ena
where $f_A(u)$ is an arbitrary function of $u$. We shall introduce similar functions
of $u$ below upon integration of the field equations. By the same token, Eqs.~\p{f4} and \p{f7} give
\bea
\label{r2}
C' e^U r^{ q+1} = \frac{ q}{16}cE + f_C(u), \\
\label{r3}
\phi' e^U r^{ q+1} = \frac{p-3}{4}\e cE + f_\phi(u).
\ena
Also from Eqs.~\p{f2}, \p{f4} and \p{f6}, we get
\bea
(r^{2 q+1} e^U U')'=0.
\ena
Thus
\bea
(e^U)' = \frac{f_U(u)}{r^{2 q+1}}.
\label{r4}
\ena
Eq.~\p{f6} can then be rewritten as
\bea
(B' e^Ur^{ q+1})'+\frac{f_U(u)}{r^{ q+1}}=-\frac{p+1}{16}cE',
\ena
which yields
\bea
B' e^Ur^{ q+1}= - \frac{p+1}{16}cE+\frac{1}{ q}\frac{f_U(u)}{r^{ q}}+f_B(u),
\label{r5}
\ena
Using the definition of $U$, Eqs.~\p{r1}, \p{r2} and \p{r5}, we find
\bea
2f_A(u)+(p-1) f_C(u)+ q f_B(u)=0.
\label{r6}
\ena
At this point, Eqs.~(\ref{r1}), (\ref{r2}), (\ref{r3}) and (\ref{r5}) have to be substituted
into (\ref{f5}), which will result in a constraint on the functions of $u$ introduced
thus far. Thereafter, one has to analyze (\ref{f3}), which will give ordinary
differential equations for the non-linear wave amplitudes, and (\ref{f1})
to determine $K$. By use of Eq.~(\ref{f2}), (\ref{f1}) can actually be simplified to
\begin{equation}
2(p-1) (\ddot{C} - 2 \dot A\dot{C} +\dot{C}^2) +2( q+2)( \ddot B - 2
\dot A \dot B + \dot B^2)+\dot{\phi}^2= e^{2(A-B)}\frac{(e^Ur^{ q+1}K')'}{e^Ur^{ q+1}}
+e^{2(A-C)}\partial_\alpha^2 K\, .
\label{Ksimple}
\end{equation}
It is convenient to enforce the Brinkmann parametrization (\ref{brink}) for the plane
wave at $r\to\infty$.
Then $A$, $B$ and $C$ go to 0 for large $r$.
For the asymptotic large $r$ plane wave, it is not difficult to include
the $\alpha\beta$ polarizations by assuming
\begin{equation}
K(u,r,y^\alpha)=k(u,r)+K_{\alpha\beta}(u)y^\alpha y^\beta.
\label{Kkr}
\end{equation}
Then, Eq.~\p{Ksimple} reduces to
\begin{equation}
2(p-1) (\ddot{C} - 2 \dot A\dot{C} +\dot{C}^2) +2( q+2)( \ddot B - 2
\dot A \dot B + \dot B^2)+\dot{\phi}^2= e^{2(A-B)}\frac{(e^Ur^{ q+1}k')'}{e^Ur^{ q+1}}
+e^{2(A-C)}K_{\alpha\alpha}.
\label{krsimple}
\end{equation}
One then simply takes a solution for $A$, $B$, $C$ and $\phi$ (with all the arbitrary
functions of $u$ contained therein), specifies $K_{\alpha\beta}(u)$ and solves the above
equation for $k(r)$. The arbitrary functions of $u$ contained in our solution are
thus $K_{\alpha\beta}(u)$ and the functions of $u$ contained in the solution of
(\ref{f3})--(\ref{f8}) subject to $A$, $B$ and $C$ going to 0 at large $r$.

Even though the equation for $k$ cannot be solved in terms of elementary functions
due to the complexity of $r$-dependences in $A$, $B$, $C$, $\phi$ (exact expressions
will be given in the subsequent sections), $k$ can be always expressed as
two integrations over $r$ of a given $r$- and $u$-dependent quantity.
(We could set $k=0$, but then this would give a constraint on the solution~\cite{MOTW2}.)

Our final step is to give an explicit analysis of (\ref{f3})--(\ref{f8}).
Since the general supergravity case is quite complicated algebraically, we shall
first set $c$ to 0 (i.e., assume no $p$-form charge) and see how this type of derivations
works (there is a rather non-trivial interplay between the structure of the constraint
equation (\ref{f5}) and the amplitude evolution equation (\ref{f3})).
We shall then proceed with the case of finite charge non-extremal supergravity $p$-branes.

\section{Pure and dilaton gravity}

In view of its algebraic simplicity, it seems reasonable to consider the case of vanishing form charge first.
This amounts to simply setting $c=0$ in the equations of the previous section.

One can use (\ref{f6}) and (\ref{r4}) to eliminate second derivatives from (\ref{f5}):
\bea
-U'^2-\frac{2( q+1)}{r}U'-(2A'+(p-1)C'+U')B'+2A'^2+(p-1)C'^2+\frac12 \phi'^2=0,
\ena
or, expressing $B$ through $U$, $A$ and $C$:
\bea
2A'^2+(p-1)C'^2+\frac{(2A'+(p-1)C')^2}{ q}+\frac12 \phi'^2
-\frac{U'^2( q+1)}{ q}-\frac{2( q+1)U'}{r} = 0.
\label{constr1}
\ena
We then multiply this equation by $r^{2( q+1)}e^{2U}$ and use (\ref{r1}), (\ref{r2})
and (\ref{r4}) together with
\bea
e^U=h_{U}-\frac{f_U}{2 q r^{2 q}}.
\label{eU}
\ena
The $r$-dependences in the last two terms of (\ref{constr1}) cancel each other,
leaving an $r$-independent constraint equation:
\bea
2f_A^2+(p-1)f_C^2+\frac{(2f_A+(p-1)f_C)^2}{ q}-2( q+1)f_Uh_U+\frac{1}{2}f_\phi^2=0,
\label{fuhu}
\ena
or
\bea
( q+2)f_A^2+2(p-1)f_Af_C+\frac{(p-1)(D-4)}{2}f_C^2+\frac{ q}{4}f_\phi^2- q ( q+1)f_Uh_U=0.
\label{constr}
\ena
(Note that, from (\ref{fuhu}), $f_Uh_U$ is a sum of squares, and thus positive.)

We now turn to Eq.~(\ref{f3}). Expressing $B$ through $U$, $A$ and $C$, we obtain
\bea
-( q+2)(\dot A'+A'\dot U)-(p-1)(\dot C'+C'\dot U)+2( q+2)A'\dot A
+2(p-1)(A'\dot C+C'\dot A)\nn
+\frac{ q}{2}\dot\phi \phi'+(p-1)(D-4)C'\dot C +( q+1)\dot U'=0 .
\ena
We then multiply this equation by $r^{ q+1}e^U$ and use (\ref{r1}, \ref{r2}) together
with their $u$-derivatives. The result is
\bea
-( q+2)\dot f_A -(p-1)\dot f_C+2( q+2)f_A\dot A+2(p-1)(f_A\dot C+f_C\dot A)+\frac{ q}{2}f_\phi\dot\phi \nn
+(p-1)(D-4)f_C\dot C +( q+1)r^{ q+1}e^U\dot U'=0.
\label{ACU}
\ena
We then integrate (\ref{r1}, \ref{r2}) to obtain
\bea
A=h_A(u) -\frac{\sqrt{2}f_A}{\sqrt{ q h_Uf_U}}\,\mbox{arctanh}
\left(\sqrt\frac{f_U}{2 q h_U}\frac1{r^q}\right),
\ena
where $h_A(u)$ is an arbitrary function of $u$, and an analogous expression
for $C$ and $\phi$. Differentiating these formulas with respect to $u$ yields
\bea
\dot A = \dot h_A-\left(\frac{\sqrt{2}f_A}{\sqrt{ q h_Uf_U}}\right)^\textbf{.}\mbox{arctanh}
\left(\sqrt\frac{f_U}{2  q h_U}\frac1{r^{ q}}\right) \nn
-\frac{f_A}{ q}\sqrt\frac{h_U}{f_U}\left(\sqrt\frac{f_U}{h_U}\right)^\textbf{.}
\frac{e^{-U}}{r^{ q}},
\ena
and an analogous expression for $\dot C$ and $\dot \phi$.
Note that $\mbox{arctanh}\, x = \frac12 \ln\frac{1+x}{1-x}$, and this looks similar to
the solution in \cite{intersect}.
Furthermore, using (\ref{eU}), we obtain
\bea
\dot U'= ((e^U)'/e^U)^\textbf{.}=-\frac{f_U^2}{r^{2 q+1}e^{2U}}
\left(\frac{h_U}{f_U}\right)^\textbf{.}.
\ena
Substituting the above expressions for time derivatives into (\ref{ACU}), we find
that there are only three different dependences on $r$ present in that equation.
The terms independent of $r$ are
\bea
M \equiv -( q+2)\dot f_A -(p-1)\dot f_C+2( q+2)f_A\dot h_A+2(p-1)(f_A\dot h_C+f_C\dot h_A)
\nn
+(p-1)(D-4)f_C\dot h_C+\frac{ q}{2} f_\phi \dot h_\phi.
\label{res1}
\ena
The terms proportional to
$\mbox{arctanh}\left(\sqrt\frac{f_U}{2 q h_U}\frac1{r^q}\right)$ are
\bea
-\sqrt\frac{h_Uf_U}{ q}\left(\frac{( q+2)f_A^2+2(p-1)f_Af_C+\frac{(p-1)(D-4)}{2}f_C^2
+ \frac{ q}{4} f_\phi^2}{h_Uf_U} \right)^\textbf{.} \nn
\times \mbox{arctanh}\left(\sqrt\frac{f_U}{2 q h_U}\frac1{r^q}\right),
\ena
which vanishes by (\ref{constr}). The terms proportional to $\frac{e^{-U}}{r^{ q}}$ are
\bea
-\Bigg[\frac1{ q}\sqrt\frac{h_U}{f_U}\left(\sqrt\frac{f_U}{h_U}\right)^\textbf{.}
\left(2( q+2)f_A^2+4(p-1)f_Af_C+(p-1)(D-4)f_C^2+\frac{ q}{2} f_\phi^2\right) \nn
+( q+1)f_U^2 \left(\frac{h_U}{f_U}\right)^\textbf{.}\Bigg]\frac{e^{-U}}{r^{ q}},
\ena
which again vanishes by (\ref{constr}).

We have therefore shown that (\ref{f3}) reduces to a single ordinary differential equation
with respect to $u$ (since all the terms except for (\ref{res1}) vanish
when (\ref{constr}) is satisfied):
\bea
M=0.
\label{dotprimeconstr}
\ena

As per discussion under (\ref{Ksimple}), it is convenient to enforce the Brinkmann
parametrization for the plane wave at large $r$, in which case $A$, $B$ and $C$ should
go to 0 for $r$ going to infinity. This translates into $h_A=h_C=0$, $h_U=1$.
We are then left with five functions ($f_A$, $f_C$, $f_U$, $h_\phi$, $f_\phi$) subject
to two constraints (\ref{constr}) and (\ref{dotprimeconstr}), and additionally
$K_{\alpha\beta}(u)$, which are arbitrary functions of $u$, as explained under
Eq.~(\ref{krsimple}). Of these, $K_{\alpha\beta}(u)$ and $h_\phi$ represent space-filling
gravitational waves (whose large $r$ asymptotics in the Brinkmann form flat-space
dilaton-gravity plane wave) and the remaining functions do not contribute to
the large $r$ behavior of the metric (and hence represent non-linear waves localized
on the extended black hole worldvolume).

For the vanishing $p$-form charge case (which is what we are considering in this section),
the dilaton can be consistently set to zero. In this case, we end up with pure gravity
(empty space) solutions. It is interesting that, for the pure gravity case, the
space-filling and worldvolume modes completely decouple. Indeed, the only space-filling modes
in that case are $K_{\alpha\beta}(u)$ (which are arbitrary functions, not affected
by the values of $f_A$, $f_C$, $f_U$). On the other hand, $K_{\alpha\beta}(u)$ do not
appear in equations (\ref{constr}) and (\ref{dotprimeconstr}) constraining the behavior of
the non-linear modes localized on the worldvolume. (In particular, there are always
solutions for which the only $u$-dependence is contained in the $uu$-component of the
metric of our ansatz, corresponding to zero amplitudes of the worldvolume modes.)
There could be deeper symmetry reasons for this surprising
decoupling of non-linear modes.

We conclude this section by a summary of the expressions for our solution in a more
compact notation. We introduce $R=(f_U/2q)^{1/2q}$, $\alpha=f_A/\sqrt{2qf_U}$, $\gamma=f_C/\sqrt{2qf_U}$, $\omega=f_\phi/\sqrt{2qf_U}$ and rename $h_\phi$ to $h$. Then the solution is given by (\ref{met}) with
\bea
e^A&=&\left(\frac{1-(R(u)/r)^q}{1+(R(u)/r)^q}\right)^{\alpha(u)},\\
e^C&=&\left(\frac{1-(R(u)/r)^q}{1+(R(u)/r)^q}\right)^{\gamma(u)},\\
e^\phi&=&e^{h(u)}\left(\frac{1-(R(u)/r)^q}{1+(R(u)/r)^q}\right)^{\omega(u)},\\
e^B&\equiv& e^{U/q}  e^{-2A/q-(p-1)C/q}=\left(1-\left(\frac{R(u)}{r}\right)^{2q}\right)^{1/q}\left(\frac{1-(R(u)/r)^q}{1+(R(u)/r)^q}\right)^{-(2\alpha(u)+(p-1)\gamma(u))/q}\nn
&=&\frac{\left({1-(R(u)/r)^q}\right)^{-(2\alpha(u)+(p-1)\gamma(u)-1)/q}}{\left(1+(R(u)/r)^q\right)^{-(2\alpha(u)+(p-1)\gamma(u)+1)/q}}.
\ena
The functions of $u$ appearing above satisfy
\bea
2( q+2)\alpha^2+4(p-1)\alpha\gamma+(p-1)(D-4)\gamma^2+\frac{ q}{2}\omega^2&=&q+1,\\
-(q+2)R^{-q}(\alpha R^q)^\textbf{.}-(p-1)R^{-q}(\gamma R^q)^\textbf{.}+\frac{q}2 \omega \dot h&=&0,
\ena
which follow from (\ref{constr}), (\ref{res1}) and (\ref{dotprimeconstr}).
$K$ is then determined by (\ref{Kkr}) and (\ref{krsimple}).

\section{Supergravity}

Having examined the simple case of pure and dilaton gravity, we now turn to
the general case with nonvanishing field strength. This set-up can be naturally implemented in ten-dimensional
supergravity.

Substituting Eqs.~\p{r1}, \p{r2}, \p{r3} and \p{r5} into \p{f5}, we get
\bea
\label{doubleprime}
( q+2)f_A^2+2(p-1)f_Af_C+\frac{(p-1)(D-4)}{2}f_C^2
+\frac{ q}{4} f_\phi^2- q ( q+1)f_Uh_U \nn
+\frac{ q}{4}\Big{\{}(c E)^2 +2\Big[
2f_A+(p-1)f_C+\frac{p-3}{4}\e f_\phi \Big] cE - e^U r^{ q+1}
cE'\Big{\}}=0.
\ena

Let us solve Eq.~(\ref{doubleprime}), which can be written as
\begin{equation}
\label{Riccati} c E'= Q \Big[ (c E)^2+ 2 \kappa c E - \sigma^2\Big]\, ,
\end{equation}
where we have introduced
\begin{eqnarray}
Q &=& e^{-U} r^{-( q+1)}, \nn
\sigma^2&=&-\frac{4}{ q}\Big{\{}f_A^2( q+2)+2(p-1)f_Af_C+\frac{(p-1)(D-4)}{2}f_C^2
+\frac{ q}{4} f_\phi^2- q ( q+1)f_Uh_U\Big{\}}\, .
\end{eqnarray}
Eq.~(\ref{Riccati}) is a special Riccati
equation for $cE$, and it obviously has a particular
solution:\footnote{For a general Ricatti equation $g(x)y'=f_2(x)
y^2+f_1(x) y +f_0(x)$, the general solution is given by
$$
y(x)=y_0(x) + \Phi(x)\Big[\mathrm{const}-\int
\Phi(x)\frac{f_2(x)}{g(x)}dx\Big]^{-1}\, ,
$$
where $y_0$ is a particular solution and
$$
\Phi(x)=\exp{\Bigg{\{}\int \Big[2f_2(x)y_0(x)+
f_1(x)\Big]\frac{1}{g(x)}dx\Bigg{\}}}\, .
$$
In our case, $\Phi=e^{2\tau w}$. We get the particular solution
when the integral constant $``\mathrm{const}"$ approaches infinity.
In our present context, this integration constant can be an arbitrary
function of $u$.}
\begin{equation}
c E_0= -\kappa + \tau \, ,
\end{equation}
where
\bea
\kappa &=& 2f_A + (p-1)f_C +\epsilon\Big(\frac{p-3}{4}\Big)f_{\phi}\,, \nn
\tau^2 &=& \kappa^2+\sigma^2.
\label{kt}
\ena
So the general solution of the Riccati Eq.~(\ref{Riccati}) is given by
\bea
cE &=& cE_0 + e^{2\tau w }\Bigg(h_E- \int e^{2\tau w } Q dr\Bigg)^{-1} \nn
&=& -\kappa + \tau + \frac{ 2\tau e^{2\tau w}}{2\tau h_E+1- e^{2\tau w}}\, ,
\label{Esolution1}
\ena
where $h_E$ is an arbitrary function of $u$, as explained in the preceding footnote.
Note that Eq.~\p{f5} gave a constraint~\p{dotprimeconstr} on the $u$-dependent functions
when we set the field strength to zero, but here it determines the field strength
instead of constraining the functions.

With the result~(\ref{Esolution1}), we find
\begin{equation}
\label{z}
z \equiv \int cE Q dr=(-\kappa
+\tau)w-\ln{\Big{|}\frac{1}{2\tau}\Big(2\tau h_E+1-e^{2\tau w}\Big)\Big{|}}\, .
\end{equation}
After substituting (\ref{z}) into Eqs.~(\ref{r1}-\ref{r3}),
we get the expressions for $A$, $C$ and
$\phi$:
\begin{eqnarray}
\label{ACPHI1}
&&A=h_A+ f_A w+ \frac{ q}{16}\Big{\{}(-\kappa
+\tau)w-\ln{\Big{|}\frac{1}{2\tau}\Big(2\tau h_E+1-e^{2\tau w}\Big)\Big{|}}\Big{\}}\, ,\nn
&&C=h_C+ f_C w+ \frac{ q}{16}\Big{\{}(-\kappa
+\tau)w-\ln{\Big{|}\frac{1}{2\tau}\Big(2\tau h_E+1-e^{2\tau w}\Big)\Big{|}}\Big{\}}\, ,\nn
&&\phi=h_{\phi}+ f_{\phi} w+\epsilon \frac{p-3}{4}\Big{\{}(-\kappa
+\tau)w-\ln{\Big{|}\frac{1}{2\tau}\Big(2\tau h_E+1-e^{2\tau w}\Big)\Big{|}}\Big{\}}\, ,
\end{eqnarray}
where
\bea
w \equiv \int Qdr
=-\sqrt{\frac{2}{ q f_U h_U}}\,\mbox{arctanh}
\left(\sqrt\frac{f_U}{2 q h_U}\frac1{r^q}\right)\, .
\label{w}
\ena
Then Eq.~(\ref{QSE'constant}) gives
\begin{equation}
\label{NandhE}
c^2 e^{2N}=2\tau h_E + 1 \, ,
\end{equation}
where we have defined
\begin{eqnarray}
N \equiv 2 h_A + (p-1) h_C+\frac{p-3}{4}\e h_{\phi}\, .
\label{defn}
\end{eqnarray}
The above equation gives the relation between $N$ and $h_E$.

Now let us consider the dot-prime equation (\ref{f3}). After substituting
\begin{eqnarray}
\dot{A}=\dot{h}_A+ \dot{f}_A w + f_A \dot{w}+
\Big(\frac{ q}{16}\Big)\dot{z},
\end{eqnarray}
and similar expressions for $\dot{C}$ and $\dot{\phi}$ into Eq.~(\ref{f3}), we get
\begin{eqnarray}
&&M+( q+1) [r^{ q}e^U\dot U'+  q(f_Uh_U)^{\dot{~}}w+2
 q(f_Uh_U)\dot{w}] \nn
&&- \frac{ q}{2} \{ c\dot{E} - [(\kappa w)^{\cdot}+\dot{N}+\dot{z}] cE
- \kappa\dot{z}+\sigma^2 \dot{w}+\sigma\dot{\sigma}w \}=0\, ,
\label{dp1}
\end{eqnarray}
where $M$ and $N$ are defined in Eqs.~\p{res1} and \p{defn}, respectively.
With the help of Eq.~(\ref{w}), Eq.~\p{dp1} reduces to
\begin{eqnarray}
\label{dotprime2}
&&M- \frac{ q}{2} \{ c\dot{E} - ((\kappa w)^{\cdot}+\dot{N}+\dot{z}) cE
-\kappa\dot{z}+\sigma^2 \dot{w}+\sigma\dot{\sigma}w \}=0\, .
\end{eqnarray}
We also have
\begin{eqnarray}
c\dot{E} &=& \dot{\tau}-\dot{\kappa}+(cE+\kappa
-\tau)\Bigg[\frac{\dot{\tau}}{\tau}+(2\tau w)^{\cdot}-\frac{( 2\tau
h_E+1- e^{2\tau w})^{\cdot}}{( 2\tau h_E+1- e^{2\tau w})}\Bigg]\,, \nn
\dot{z} &=& (\dot{\tau}-\dot{\kappa})w+(\tau-\kappa)\dot{w}+\frac{\dot{\tau}}{\tau}-\frac{(
2\tau h_E+1- e^{2\tau w})^{\cdot}}{( 2\tau h_E+1- e^{2\tau w})}\,,
\label{hrel1}
\end{eqnarray}
and
\begin{eqnarray}
\frac{(2\tau h_E+1- e^{2\tau w})^{\cdot}}{ 2\tau h_E+1- e^{2\tau w}}
=\frac{ (2\tau h_E)^{\cdot}-(e^{2\tau w})^{\cdot}}{2\tau h_E+1- e^{2\tau w}}
=\frac{ 2(2\tau h_E+1)\dot{N}-(e^{2\tau w})(2\tau w)^{\cdot}}{2\tau h_E+1- e^{2\tau w}}\, .
\label{hrel2}
\end{eqnarray}
Here we have used the relation~\p{NandhE}. From~\p{hrel1} and \p{hrel2},
we find
\begin{eqnarray}
c\dot{E}-[ (\kappa w)^{\cdot}+\dot{N}+\dot{z}] cE - \kappa\dot{z}
=-\dot{\kappa}+(\tau+\kappa)\dot{N}- (\sigma^2
\dot{w}+\sigma\dot{\sigma}w) \, ,
\end{eqnarray}
where we have also used $\tau^2=\kappa^2+\sigma^2$ and
$\tau\dot{\tau}=\kappa\dot{\kappa}+\sigma\dot{\sigma}$. Substituting
this result into Eq.~(\ref{dotprime2}), we obtain our final form of the constraint~\p{f3}:
\begin{equation}
\label{reduceddotprime}
M+ \frac{ q}{2}\{ \dot{\kappa}-(\tau+\kappa)\dot{N}\}=0\,.
\label{ftp}
\end{equation}

To summarize, our solutions are given by the metric~\p{met} with
\bea
e^{2A} &=& e^{2h_A+(f_A+\frac{ q}{16}(\tau-\kappa))w}
\left(\frac{2\tau}{2\tau h_E +1-e^{2\tau w}} \right)^{ q/16}, \nn
e^{2C} &=& e^{2h_C+(f_C+\frac{ q}{16}(\tau-\kappa))w}
\left(\frac{2\tau}{2\tau h_E +1-e^{2\tau w}}\right)^{ q/16}, \nn
e^{2B} &=& e^{(2/ q) (U-2A-(p-1)C)}, \nn
e^{2\phi} &=& e^{2h_\phi+(f_\phi+\e\frac{p-3}{4}(\tau-\kappa))w}
\left(\frac{2\tau}{2\tau h_E +1-e^{2\tau w}} \right)^{ q/16},
\ena
where $e^U, \kappa$ and $\tau$ are given in \p{eU} and \p{kt}, and
\bea
e^w = \left(\frac{g_-}{g_+}\right)^{1/\sqrt{2 q f_U h_U}},\quad
g_\pm \equiv 1\pm \sqrt{\frac{f_U}{2 q h_U}}\frac{1}{r^q},
\ena
and use has been made of the relation
${\rm arctanh} x=\frac12 \ln\left(\frac{1+x}{1-x}\right)$.
The functions $g_\pm$ are time-dependent harmonic functions.
The $u$-dependent functions $h_A, h_C, h_\phi, f_A, f_C, f_\phi$ and $h_E$
are constrained by \p{NandhE} and \p{ftp}.

In conclusion, we consider a few special cases representing dynamically invariant
sub-spaces in the space of non-linear gravitational wave modes.

\subsection{$\sigma=0$ and $\kappa\ne 0$}

In this case, we have $\tau-\kappa=0$, so the solution $A$, $C$ and
$\phi$ becomes
\begin{eqnarray}
\label{ACPHI2} &&A=h_A+ f_A w- \frac{ q}{16} \ln{\Big{|}\frac{1}{2\kappa}\Big(2\kappa
h_E+1-e^{2\kappa w}\Big)\Big{|}} \, ,\nonumber\\
&&C=h_C+ f_C w- \frac{ q}{16} \ln{\Big{|}\frac{1}{2\kappa}\Big(2\kappa
h_E+1-e^{2\kappa w}\Big)\Big{|}} \, ,\nonumber\\
&&\phi=h_{\phi}+ f_{\phi} w-\epsilon \frac{p-3}{4} \ln{\Big{|}\frac{1}{2\kappa}\Big(2\kappa
h_E+1-e^{2\kappa w}\Big)\Big{|}} \, .
\end{eqnarray}
Eq.~(\ref{reduceddotprime}) becomes
\begin{equation}
M + \frac{ q}{2} [\dot{\kappa}-2\kappa\dot{N}]=0\, .
\end{equation}

\subsection{$\kappa=0$ and $\sigma\ne 0$}

In this case we have $\tau=\sigma$, and the functions $A$, $C$ and
$\phi$ are given by
\begin{eqnarray}
&&A=h_A+ f_A w+\frac{ q}{16}\Big{\{}\sigma
w-\ln{\Big{|}\frac{1}{2\sigma}\Big(2\sigma h_E+1-e^{2\sigma
w}\Big)\Big{|}}\Big{\}}\, ,\nonumber\\
&&C=h_C+ f_C w+ \frac{ q}{16}\Big{\{}\sigma
w+\ln{\Big{|}\frac{1}{2\sigma}\Big(2\sigma h_E+1-e^{2\sigma
w}\Big)\Big{|}}\Big{\}}\, ,\nonumber\\
&&\phi=h_{\phi}+ f_{\phi} w+ \epsilon \frac{p-3}{4}\Big{\{}\sigma
w+\ln{\Big{|}\frac{1}{2\sigma}\Big(2\sigma h_E+1-e^{2\sigma
w}\Big)\Big{|}}\Big{\}}\, .
\end{eqnarray}
Eq.~(\ref{reduceddotprime}) becomes
\begin{equation}
M- \frac{ q}{2} \sigma\dot{N} =0\, .
\end{equation}

\subsection{$\tau=0$}

In the $\tau\rightarrow 0$ limit, the solution (\ref{ACPHI1}) reduces to
\begin{eqnarray}
\label{ACPHI3} &&A=h_A+ f_A w-
\frac{ q}{16}\Big{\{}\kappa w+\ln{|h_E-w|}\Big{\}}\, ,\nonumber\\
&&C=h_C+ f_C w-\frac{ q}{16} \Big{\{}\kappa w + \ln{|h_E-w|}\Big{\}}\, ,\nonumber\\
&&\phi=h_{\phi}+ f_{\phi} w-\epsilon \frac{p-3}{4}\Big{\{}\kappa w +\ln{|h_E-w|}\Big{\}}\, .
\end{eqnarray}
Eq.~(\ref{NandhE}) does not give any constraint on $h_E$.
The dot-prime equation reduces to
\begin{equation}
M+\frac{ q}{2}\dot{\kappa}=0\, .
\end{equation}

\subsection{$\kappa= 0$ and $\sigma=0$}

This special case is very simple.
The function $A$, $C$ and $\phi$ are given by
\begin{eqnarray}
\label{ACPHI4}
&& A=h_A+ f_A w-\frac{ q}{16} \ln{|h_E-w|} \, ,\nonumber\\
&&C=h_C+ f_C w- \frac{ q}{16} \ln{|h_E-w|} \, ,\nonumber\\
&&\phi=h_{\phi}+ f_{\phi} w-\epsilon \frac{p-3}{4} \ln{|h_E-w|} \, .
\end{eqnarray}
The dot-prime equation is also simple, which is just $M=0$.
Thus the functions
$f_A$, $f_C$, $f_{\phi}$, $h_A$, $h_C$, $h_{\phi}$, $h_E$, $f_U$ and
$h_U$ are constrained by
\begin{equation}
\sigma=\kappa=N=M=0\, .
\end{equation}
If we also require the solution approaches a Brinkmann form plane wave when
$r\rightarrow \infty$, we have to require the functions in $A$ and
$C$ satisfy additional two relations. So, there are totally six
constraints on these functions, and most of them are not
independent. In this setting, it is possible to prove that $f_Uh_U>0$
(which is something we have implicitly assumed).

Let us consider a simple example (specifying $D=10$): We set $h_E=1$, and
$e^A\rightarrow 1$ and $e^C\rightarrow 1$ gives $h_A=h_C=0$. The condition
$M=0$ reduces to
\begin{eqnarray}
0=-( q+2)\dot f_A -(p-1)\dot
f_C+\frac{ q}{2}f_{\phi}\dot{h}_{\phi}\, .
\label{m0}
\end{eqnarray}
In addition, $f_{\phi}$ and $h_{\phi}$ have to satisfy $\kappa=N=0$,
i.e.,
\begin{eqnarray}
\label{sk1}
&&0=2f_A + (p-1)f_C +\epsilon \frac{p-3}{4}f_{\phi}\,
,\\
&&0=\epsilon \frac{p-3}{4} h_{\phi}\, .
\label{sk2}
\end{eqnarray}
Further, $\sigma=0$ gives
\begin{eqnarray}
0=f_A^2( q+2)+2(p-1)f_Af_C+3(p-1)f_C^2 +\frac{ q}{4}
f_\phi^2- q ( q+1)f_Uh_U\, .
\label{constf}
\end{eqnarray}
We now consider all possible cases separately.
\begin{enumerate}
\item[(i)]
For $p=3\; ( q=4)$, we get $f_A=-f_C$ from \p{sk1}, and Eq.~\p{constf} gives
\begin{equation}
8f_A^2+f_{\phi}^2=20f_Uh_U>0\, .
\end{equation}
\item[(ii)]
For $p=1\; ( q=6)$, we have $h_\phi=0$ from \p{sk2} and then $f_A$ is a constant from \p{m0}.
Then $f_{\phi}^2=16f_A^2$ from \p{sk1} and Eq.~\p{constf} gives
\begin{equation}
16f_A^2=21f_Uh_U>0 \, .
\end{equation}
\item[(iii)]
If $p\ne 3$ and $p\ne 1$, we get $h_{\phi}=0$, and
\begin{equation}
s=( q+2)f_A +(p-1)f_C\, ,
\end{equation}
is a constant. Substituting this into Eq.~\p{constf}, we get the second order equation
\begin{equation}
128f_A^2-16(p+1)f_As+\frac{16p-(p+1)^2}{ q}s^2
=(p-3)^2(p-1)( q+1)f_Uh_U\, .
\label{80}
\end{equation}
Since its discriminant is always negative:
\begin{equation}
\Big(16(p+1)f_A\Big)^2-4\times
128f_A^2\times\Bigg(\frac{16p-(p+1)^2}{ q}\Bigg) =
-f_A^2(p-3)^2(p-1)<0\, ,
\end{equation}
and $16p-(p+1)^2>0$ for $9\geq p \geq 1$, we always have the left-hand side of Eq.~\p{80}
positive
for arbitrary $s$.
This means that $f_Uh_U>0$.
\end{enumerate}
Thus in all cases, we have $f_Uh_U>0$ as we have mentioned before.

\section{Conclusions}

In this paper, starting with an action for
form fields and dilaton coupled to gravity, we have derived general solutions describing extended higher-dimensional black holes with strong
dilaton-gravity plane waves propagating along their worldvolume. Both space-filling and (localized) worldvolume non-linear wave modes are present in our solutions,
giving a large number of arbitrary functions of the light-cone time (corresponding
to the non-linear wave amplitudes).

When the field strength is set to zero, the dilaton can also be set to zero and the
solution becomes that in pure gravity. It turns out that the space-filling and
worldvolume non-linear gravitational wave modes are completely decoupled from each other in that case. We have also obtained solutions with a form field and nontrivial dilaton in a class of theories including ten-dimensional supergravities, and discussed their special cases.

We shall conclude with a few comments on the novelty and possible applications of our solutions.
It is quite remarkable that interactions of an extended black hole with a strong
gravitational wave can be described exactly, and one might be wondering for
the underlying reason that makes the equations solvable. When supersymmetric solutions
of \cite{CDEG,MOTW2} were presented, one could be suspecting that supersymmetry
linearizes (trivializes) the gravitational dynamics in some sense.
We have presently constructed solutions (involving gravitational waves of an arbitrary
profile) that break supersymmetry completely. Hence, integrability is not related
to supersymmetry in this case, and must have roots in other features of our set-up,
most likely, the light-like isometry. The structure of the equations we find appears
quite general, and it is shared by a number of different cases (extremal and non-extremal
black holes, single and intersecting $p$-branes). Indeed, we plan to report~\cite{CEO} on non-extremal intersecting brane solutions analogous to the single black $p$-brane solutions
we have described here. For the case of pure gravity, we have presented simple yet non-trivial solutions that appear to have been overlooked in the long development of that subject (perhaps, because they only exist in more than four space-time dimensions). We have also presented, for the first time to the best of our knowledge, exact solutions involving non-linear oscillations localized in the near-horizon region of extended black holes
(solutions of \cite{CDEG,MOTW2} only involved non-linear waves propagating through the entire space-time).

We firmly believe that exact non-linear solutions are of interest in their own right (one can think of the broad and fruitful investigations of black rings and related solutions in higher dimensions, see, e.g., \cite{emparan}). However, it is worth mentioning that similar set-ups (an extended black object with a wave propagating along its worldvolume) have appeared in a number of string-theoretical considerations.
Thus, light-like `cosmologies' of \cite{lightlike} start with a brane embedded
into a singular gravitational wave (the actual papers consider a restricted class
of wave profiles, since our general solutions were not available at the time).
This system is then used to formulate a time-depended analog of the AdS/CFT
correspondence, which relates an AdS-like space-time with a light-like singularity
to a gauge theory with light-like time dependences inserted in its Lagrangian.
The kind of solutions we have presently developed suggests various generalizations
of this construction. Furthermore, oscillations localized on the worldvolume of black
objects have appeared in some string-theoretic approaches to Hawking radiation
\cite{cm}. In the D-brane language, these oscillations are represented by open
strings moving along the branes. In the smooth space-time language, the oscillations
are nothing but the near-horizon gravitational waves, for which we have
presently given a full non-linear description. Constructing intersecting brane
generalizations of our solutions will take us even closer to the situation
considered in \cite{cm}.

\section*{Acknowledgement}

This work was supported in part by the Grant-in-Aid for
Scientific Research Fund of the JSPS (C) No. 20540283, No. 21$\cdot$09225 and
(A) No. 22244030. The work of O.E. has been supported by grants from the Chinese Academy
of Sciences and National Natural Science Foundation of China.


\end{document}